\input phyzzx
\def\e{\adveq\eqno{\rm (\chapterlabel\the\equanumber)}}

\def\adveq{\global\advance\equanumber by 1}
\def\myeq{{\rm \chapterlabel\the\equanumber}}

\def\semidirect{\mathrel{\raise0.04cm\hbox{${\scriptscriptstyle |\!}$
\hskip-0.175cm}\times}}

\def\mod{\mathop{\rm mod}\nolimits}

\def\ref#1{$^{[#1]}$}

\def\r#1{$[\rm#1]$} 
\def\twidle{\tilde}

\def\e{\adveq\eqno{\rm (\chapterlabel\the\equanumber)}}

\def\adveq{\global\advance\equanumber by 1}
\def\myeq{{\rm \chapterlabel\the\equanumber}}

\def\semidirect{\mathrel{\raise0.04cm\hbox{${\scriptscriptstyle |\!}$
\hskip-0.175cm}\times}}

\def\mod{\mathop{\rm mod}\nolimits}

\def\ref#1{$^{[#1]}$}

\def\r#1{$[\rm#1]$} 
\def\twidle{\tilde}

%

%

\date{August, 2006}
\titlepage
\title{Galois Groups in Rational Conformal Field Theory II.}
\line{\hfil \titlestyle{The Discriminant.}\hfil}
\author{Doron Gepner}
\line{\hfill Department of Particle Physics\hfill}
\line{\hfill Weizmann Institute\hfill}
\line{\hfill Rehovot 76100, Israel\hfill}
\abstract
We express the discriminant of the polynomial relations 
of the fusion ring, in any conformal field theory, as the product
of the rows of the modular matrix to the power $-2$. The discriminant
is shown to be an integer, always, which is a product of primes which
divide the level. Detailed formulas for the discriminant are given
for all WZW conformal field theories. 
\endpage

The classification of rational conformal field theories is one  
of the intriguing problems in relation to string theory and
to critical phenomena. 
The study of such theories were initiated in the seminal works
\REF\BPZ{A.A. Belavin, A.M. Polyakov and A.B. Zamolodchikov, Nucl.
Phys. B 241 (1984) 33.}
\REF\Witten{E. Witten, Commun. Math. Phys. 92 (1984) 455.}
\REF\KZ{V. Knizhnik and A.B. Zamolodchikov, 
Nucl. Phys. B 247 (1984) 83.}
\REF\GW{D. Gepner and E. Witten, Nucl. Phys. B 278 (1986) 493.}
\r{\BPZ,\Witten,\KZ,\GW}.

Here we continue the work of ref. 
\REF\Galois{``Galois groups in rational 
conformal field theory", hep-th/0606081 (2006) June.}\r\Galois,
where some equations where derived for the fusion ring.
Our aim here is to define the discriminant of the relations of
the fusion ring. This is expressed as the product of the rows 
of the modular matrix to the powere $-2$. It is shown to be
an integer which has the striking property of being a 
product of primes dividing the level. We hope that these results
will help in the classification of rational conformal field theories.
 
From the discussion in ref. \r\Galois,
we recall the theorem proven originally in ref. \REF\SB{D. Gepner, 
Commun. Math. Phys. 141 (1991) 381.}\r\SB,
which states as follows:

{\rm \bf Theorem (1): }
Any fusion ring is a ring of polynomials $Z[x_1,x_2,\ldots, x_m]/I$
where $I$ is the ideal of all the polynomials vanishing on the points
of the fusion variety, $x^i_\alpha={S_{i\alpha}^\dagger\over S_{i0}}$,
where $\alpha=1,2,\ldots,m$, label the generators, and $i$ denotes the different
points of the variety, whose number is the number of primary fields.
Furthermore, the value of any primary field $[a]$ evaluated on any of
the points of the fusion variety is given by 
$p_a(x_i^1,x_i^2,\ldots,
x_i^m)={S_{a,i}^\dagger\over S_{0,i}}$.

For further explanation and a proof see refs. \r{\Galois,\SB}.

For simplicity, we assume that the fusion ring is generated by one primary field,
which we label as $x=[1]$. It follows that the fusion ring  
is given by the quotient,
$$R\approx {Z[x]\over (q(x))},\e$$
where $Z[x]$ is the ring of polynomials in $x$ with integer coefficients, and
$q(x)$ is the relation in this ring.
We can write the primary field in this ring as $p_r(x)=x^r+a_1 x^{r-1}+\ldots$,
where by convention $p_0(x)=1$ and $p_1(x)=x$ and $r$ goes from $0$
to $n-1$. Using the theorem we may write the following matrix equation,
$$M_{i\alpha}={S_{i\alpha}^\dagger\over S_{i0}}=\pmatrix{1&1&\ldots & 
1\cr
x_0&x_1&\ldots&x_{n-1}\cr
p_2(x_0)&p_2(x_1)&\ldots& p_2(x_{n-1})\cr
\ldots&\ldots&\ldots&\ldots\cr
p_{n-1}(x_0) & p_{n-1}(x_1) &\ldots &p_{n-1}(x_{n-1})\cr},\e$$
where $x_0,x_1,\ldots,x_{n-1}$ lable the points of the fusion variety,
i.e., the solutions of the equation $q(x)=0$.

Now, take the determinant of both sides of eq. (2). On the right side,
since we can eliminate the lower order terms by adding rows recursively,
we have the Vandermonde determinant,
$${\rm det}(M)=\det_{i,j} x_i^{j},\e$$
As is well known, this determinant is given as the product,
$${\rm det}(M)=\prod_{i<j} (x_i-x_j).\e$$
Squaring the Vandermonde determinant we get the discriminant,
$$D=\prod_{i<j}  (x_i-x_j)^2,\e$$
which since it is a symmetric polynomial in $x_i$ can be expressed in terms
of the coefficients of the polynomial $q(x)$. Say,
$$q(x)=x^m-S_1 x^{m-1}+S_2 x^{m-2}-\ldots+ (-1)^{m} S_{m},\e$$
where the $S_r$ are the symmetric polynomials of the roots,
$$S_r=\prod_{i_1<i_2<\ldots<i_r} x_{i_1} x_{i_2}\ldots x_{i_{r}}.\e$$

The discriminant may easily be computed for any equation. For example,
see the appendix of ref. \r\SB\ for a Mathematica program. We
quote the result for equations of order $2$ and $3$,
$$\eqalign{R_2&=S_1^2 -4 S_2,\cr
           R_3&=S_1^2 S_2^2-4 S_2^3-4 S_1^3 S_3+18 S_1 S_2 S_3-27 
S_3^2.\cr}\e$$
Very importantly, since the coefficients $S_i$ are integers, the discriminant is always
an integer.

Now, consider the left hand side of eq. (2). Since $S$ obeys, $S^2=C$,
where $C$ is the charge conjugation matrix, $\det(S)=i^s$, where $s$
is some integer. 
So we find for the determinant,
$$\det(M)=i^s\prod_{i} S_{i0}^{-1}.\e$$
Comparing eqs. (4,9), we find immediately an expression for the discriminant,
$$|D|=\prod_i S_{i0}^{-2}.\e$$
Since we are free to act with an element of the Galois group, $G$, we
find also that
$$|D|=\prod_{i} S_{i,a}^{-2},\e$$
where $a$ is any primary field in the path of the Galois element,
$$a=t(0),\qquad {\rm where\ } t\in G.\e$$

Quite strikingly, since the discriminant is an integer, we find 
that the product of the elements of the modular matrix is always the
inverse square root of an integer,
$$D=\prod_i S_{0,i}^{-2}={\rm integer}.\e$$

Although our proof utilized the assumption of one generator, we
believe that eq. (13), holds for any number of generators since it
depends only on the product of the rows of the modular matrix.
In particular, since the generator is not used, the discriminant is
independent of the choice of generators, and is always the same
for any generator.

Actually, we may calculate the determinant with a generator which is
not a primary field, thus proving that eq. (13) holds for any number 
of generators.

Of course, since the points $x_i$ have to be distinct, which is a consequence
of the semi--simplicity of the ring, the discriminant $D$, cannot vanish.
 
For example, take the WZW model $SU(2)_2$. Here 
the first row of the modular matrix reads,
$$S_{i0}=\left (1/2,1/\sqrt{2},1/2 \right).\e$$
and so the product of the row to the power of $-2$ is,
$$D=32.\e$$
The relation for this ring is $q(x)=x^3- 2 x$, and so from eq. (8),
we find for the discriminant,
$$R_3=-4 S_2^3 =32,\e$$
agreeing with eq. (15).

Before proceeding, it will be useful to work out some examples.
For WZW models we find the following formula to calculate the
modular matrix very useful,
$$S_{0,\lambda}=M \prod _{\alpha>0} \sin\left[ {\pi (\lambda+\rho)\alpha\over k+g}\right],\e$$
where $\lambda$ is a highest weight at the central charge $k$, 
$\rho$ is half
the sum of positive roots, $g$ is the dual Coxeter number, and
$\alpha$ runs over all the positive roots.
$M$ is some normalization independent of $\lambda$.

For $SU(2)_k$ WZW model we have for $S$,
$$S_{0,r}=\sqrt{2\over k+2} \sin\left [{\pi (r+1)\over k+2}\right],\e$$
where $r=0,1,\ldots,k$, is twice the isospin of the primary fields.
Multiplying the $S$ matrix according to eq. (13), 
we find for the discriminant,
$$D=2^{k+1} (k+2)^{k-1},\e$$
which is obviously an integer. 

As a further example take the theory $SU(2)_{2 n-1}/Z_2$. This theory
is composed from the integer isospin representations of $SU(2)$ at an
odd level \REF\Found{D. Gepner, ``Foundations of rational quantum
field theory",Caltech preprint CALT-68-1825,
hep-th/9211100, (1992).}\r\Found. Denoting
the fields by $l=1,2,\ldots,n$, we have for the modular matrix,
$$ S_{l,m}=\sqrt{4\over 2 n+1} \sin\left[{\pi (2 l-1)(2m-1)\over 2 n+1}
\right],\e$$
where $2 l-1$ is twice the isospin plus one. 
For a discussion of the Galois group of these theories see the 
appendix of ref. \r\Galois.
Multiplying the rows of the
$S$ matrix according to eq. (13), we find for the discriminant,
$$D=\prod_{l} S_{l0}^{-2}=(2 n+1)^{n-1}.\e$$
 
It will be useful to work out an example which is not related to current 
algebra. This is a theory with six primary fields, 
$\phi_1,\phi_2,\ldots,\phi_6$, where by convention $\phi_1=1$.
The fusion rules of the model are given by,
$$\eqalign{\phi_2^2&=1+2\phi_2+\phi_3+\phi_4+2 \phi_5+\phi_6\cr
           \phi_2\phi_3&=\phi_2+\phi_4+\phi_5+\phi_6\cr
           \phi_2\phi_4&=\phi_2+\phi_3+\phi_5+\phi_6\cr
           \phi_2\phi_5&=2\phi_2+\phi_3+\phi_4+\phi_5+\phi_6\cr
           \phi_2\phi_6&=\phi_2+\phi_3+\phi_4+\phi_5\cr
           \phi_3^2&=1+\phi_4+\phi_5\cr
           \phi_3\phi_4&=\phi_2+\phi_3\cr  
           \phi_3\phi_5&=\phi_2+\phi_3+\phi_5\cr
           \phi_3\phi_6&=\phi_2+\phi_6\cr
           \phi_4^2&=1+\phi_5+\phi_6\cr
           \phi_4\phi_5&=\phi_2+\phi_4+\phi_5\cr
           \phi_4\phi_6&=\phi_2+\phi_4\cr
           \phi_5^2&=1+\phi_2+\phi_3+\phi_4+\phi_5+\phi_6\cr
           \phi_5 \phi_6&=\phi_2+\phi_5+\phi_6\cr
           \phi_6^2&=1+\phi_3+\phi_5\cr}  \e$$
The first row of the modular matrix corresponding to this theory 
can be seen to be to be, to eleven digits precision,
$$S_{l,0}:= \{1/3, 1/3, 0.62646174719, 
0.51069629541, 1/3, 0.11576545177 \}.\e$$
Multiplying these numbers, we find the discriminant to be,
$$D=\prod_l S_{l0}^{-2}=531441=3^{12}.\e$$

From these examples alone we already see an interesting property of the
discriminant. Denote by $N$ the level of the conformal field theory.
This is defined as the least integer such that the dimensions
of the theory all obey,
$$\Delta N={\rm integer},\e$$
where $\Delta$ is any of the dimensions, and $N$ is the level, thus
defined.
We observe the following intriguing arithmetical result about 
the discriminant. 

{\rm \bf Theorem (2): }
The discriminant is always a product of primes 
which divide
the level,
$$|D|=\prod_a p_a^{n_a},\qquad{\rm where\ } p_a|N,\e$$
and the $n_a$ are some non-negative integer powers.

In the case of $SU(2)_k$ the level is $2 (k+2)$ and so eq. (26)
is obeyed, using eq. (19). For $SU(2)_{2 n-1}/Z_2$ the level is 
$2 n+1$ so again
eq. (26) holds, when compared with eq. (21). Finally for the non WZW example, using Vafa equations,
\REF\Vafa{C. Vafa, Phys. Lett. B 206 (1988) 421.}\r\Vafa, it can be shown that the 
level is  
$36$, and so again the arithmetic properties, eq. (26) are obeyed.

Actually, it is not hard to prove directly this result. Consider the
fusion ring $R$ over the field of integers modulo $p$, where $p$
is some prime which is strange to the level, $Z_p$. Since the
fusion coefficients are integers, we are allowed to do that.
Now, the matrix $M$ is given here by the same expression, eq. (2),
but now the coefficients are in the Galois Field of the equation
$q(x)=0$, now over $Z_p$. To get $S$ we must normalize $M$
such that each row is of length one. The only obstruction for doing so
is the coefficient in front of $S$ which is a square root of the level.
Thus, if $p$ is strange to the level, we obtain a unitary matrix $S$,
$S S^\dagger=1\mod\, p$,
and so 
$$\det S\neq0\mod p.$$
This implies that the points of the fusion variety are distinct modulo
$p$ and so the discriminant cannot vanish,
$$D\neq 0\mod p.\e$$
Thus the discriminant has to be a product of primes dividing the 
level, proving the theorem. This proof works for any number 
of generators assuming only that the discriminant is an integer.  

We turn now to the calculation of the discriminant for WZW models 
based on any group $G$. Before proceeding, we have to establish
the following result, or conjecture.

Assume some simple Lie algebra, $G$. 
Denote by $\twidle g_i$, $i=1,2,\ldots,n+1$, the dual Coxeter 
numbers of the algebra,
and by $g_i$ the Coxeter numbers. Let $m_i$, $i=1,2,\ldots,n$, be
the exponents of the algebra, where $n$ is the rank of the algebra.  
Let $M_k$ denote the number of primary fields in the Dual affine 
theory, at the level $k$.
Explicitly, $M_k$ is the number of solutions to the equation,
$$\sum_{i} n_i g_i=k,\e$$
where $n_i$ are some non--negative integers. It is clear that
a generating function for $M_k$ is
$$\sum_{k=0}^\infty  M_k z^k=\prod_{i=1}^{n+1} (1-z^{g_i})^{-1},\e$$
proven by expanding the right hand side of this equation.

The result that we need is as follows.

{\rm\bf Conjecture (1): }
For level $k$ which are not zero modulo any of the Coxeter numbers,
$g_i$, the number of primary fields in the Dual algebra,
$M_k$, is given by
$$M_k=s \prod_{i=1}^n {k+m_i\over m_i+1} ,\e$$
where $s$ is the number of primary fields at the level $k=1$,
$s=M_1$.

At the table the Coxeter numbers and the exponents of the different
Lie algebras are listed. We can check this conjecture for all the 
algebras, which are of types $A_n$, $B_n$, $C_n$, $D_n$,
$G_2$, $F_4$ and $E_r$, for $r=6,7,8$, for small levels, 
and it is always obeyed.

\overfullrule=0pt

%
%
\setbox\strutbox=\hbox{\vrule height12pt depth8pt width0pt}
\midinsert
\line{\hfill Table. \hfill}
\line{\hfill Exponents and Coxeter Numbers of simple Lie algebras.
\hfill}
\vskip15pt
\line{\hfill
\vbox{\offinterlineskip\hrule
\halign{&\vrule#&
  \strut\quad\hfil#\hfil\quad\cr
height2pt&\omit&&\omit&&\omit&&\omit &\cr
&Algebra&&Coxeter Numbers&&Dual Numbers&& Exponents&\cr
\noalign{\hrule}
height2pt&\omit&&\omit&&\omit&&\omit&\cr
&$A_l$&&$1,1,1,\ldots,1$&&$1,1,1,\ldots,1$&&$1,2,3,\ldots,l$&\cr
&$B_l$&&$1,1,2,2,\ldots,2$&&$1,1,1,2,2,\ldots,2$&&$1,3,5,\ldots,2l-1$
&\cr
&$C_l$&&$1,1,2,2,\ldots,2$&&$1,1,\ldots,1$&&$1,3,5,\ldots,2l-1$&\cr
&$D_l$&&$1,1,1,1,2,2,\ldots,2$&&$1,1,1,1,2,2,\ldots,2$&&
$1,3,5,\ldots,2 l-3,l-1$&\cr
&$E_6$&&$1,1,1,2,2,2,3$&&$1,1,1,2,2,2,3$&&$1,4,5,7,8,11$&\cr
&$E_7$&&$1,1,2,2,2,3,3,4$&&$1,1,2,2,2,3,3,4$&&$1,5,7,9,11,13,17$&\cr
&$E_8$&&$1,2,3,4,5,6,3,4,2$&&$1,2,3,4,5,6,3,4,2$&&
$1,7,11,13,17,19,23,29$&\cr
&$F_4$&&$1,2,3,4,2$&&$1,2,3,2,1$&&$1,5,7,11$&\cr
&$G_2$&&$1,2,3$&&$1,2,1$&&$1,5$&\cr
height2pt&\omit&&\omit&&\omit&&\omit&\cr
}\hrule}
\hfill}
\setbox\strutbox=\hbox{\vrule height8.5pt depth3.5pt width 0pt}
\vskip20pt\endinsert

As an example of conjecture (1) consider $A_l$ (or $SU(l+1)$). 
Substituting the exponents from the table we arrive at,
$$M_k=(k+1)(k+2)\ldots (k+l)/l!,\e$$
which can easily be verified directly by using the 
binomial theorem on eq. (29). 

Before proceeding let us collect some data about the 
simple Lie algebras. We denote my $R$ the index of the 
algebra which is defined as the number of elements,
$$R=|M/M^*|,\e$$
where $M$ is the long root lattice and $M^*$ its dual, i.e., 
the weight lattice. The index of the various algebras are,
$A_l$, $l+1$. $B_l$, $2$. $C_l$, $2$. $D_l$, $4$. $E_6$, $3$.
$E_7$, $2$. $E_8$, $1$. $F_4$, $2$. $G_2$, $3$. 

We shall need also the total Coxeter and dual Coxeter numbers, which 
are the sums of the respective Coxeter numbers,
which we denote by $h$ and $g$. For simply laced 
algebras the two numbers are the same, $g=h$, and we have,
$A_l$, $l+1$. $D_l$, $2 l-2$. $E_6$, $12$. $E_7$, $18$.
$E_8$, $30$. For the non--simply laced algebras we have:
For $B_l$, $h=2 l$, $g=2 l-1$. For $C_l$, $h=2 l$ and $g=l+1$.
For $F_4$ we have, $h=12$ and $g=9$. For $G_2$ we have, $h=6$ and
$g=4$.

The level of the WZW model based on a group $G$ at the central charge
$k$ is given by
$$N=(k+g) |M/M^*|=R(k+g).\e$$
Thus, we expect the discriminant to be a product of primes which 
divide $N$. Our result is actually stronger, giving the
exact value of the discriminant, and we will state it
in stages.    
First, we have the following,

{\rm \bf Conjecture (2a): }
The discriminant, $D$ for any WZW based on the algebra $G$,
and for the central charge $k$, such that $k+h-g\neq 0\mod \twidle g_i$,
for any $i$, is the integer
$$D=R^m (k+g)^r,\e$$
where $r$ and $m$ are integers which are expressed as  
polynomials in $k$ of
the order ${\rm rank}(G)=n$.

We shall give the exact expression for the powers $m$ and $r$.
We find it convenient to separate the simply laced case, $A$, $D$, $E$,
from the not simply laced cases, $B$, $C$, $G$, $F$. 
The result for the simply laced algebras
is:

{\rm \bf Conjecture (2b)}: For the simply laced algebras, 
and the level as in Conjecture (2a), the discriminant
is given by eq. (34), where the power $m$ is equal to the number
of primary fields at the central charge $k$, and the power 
$r$ is given by,
$$r=r_0 m (k-1)/(k+g-1),\e$$
where $r_0$ is some rational coefficient to be listed below.

Using conjecture (1) we may express the number of primary fields 
through the exponents,
$$m=c \prod{ k+m_i\over 1+m_i},\e$$
where $c$ is the number of primary fields at the central charge one.
Since $g-1$ is always
an exponent, we see that we can divide by $k+g-1$, as we do in
eq. (36).

We can list explicitly, the polynomial expressions for $r$
and $m$, for each of the simply laced algebras. These follow
by substituting the exponents from the table.

For $A_l$ at any central charge $k$, we have,
$$D=(l+1)^m (k+l+1)^r,\e$$
where
$$m=(k+1)(k+2)\ldots (k+l)/l!,\e$$
and $r$ is,
$$r=m (k-1) l/(k+l)=(k-1)(k+1)(k+2)\ldots (k+l-1)/(l-1)!.\e$$

For $D_l$ we have, for $k\neq 0\mod 2$, or for odd $k$,
$$D=4^m (k+2 l-2)^r,\e$$
where
$$m=(k+1)(k+3)(k+5)\ldots (k+2 l-3) (k+l-1)/(2^{l-3} l!).\e$$
For $r$ we have,
$$r=(k-1) l m/(k+2 l-3)=(k-1)(k+1)(k+3)\ldots (k+2 l-5)
(k+l-1)/[2^{l-3} (l-1)!].\e$$

For $E_6$ we find for the central charge $k$,
which is not divisible by $2$ or $3$,
$$D=3^m (k+12)^r,\e$$
where 
$$m=(k+1)(k+4)(k+5)(k+7)(k+8)(k+11)\cdot 3/(2\cdot 5\cdot 6\cdot 8\cdot 9
\cdot 12),\e$$
is the number of primary fields, and the power $r$ is 
$$r=(k-1)(k+1)(k+4)(k+5)(k+7)(k+8)\cdot 3/(4\cdot5\cdot 6\cdot 8\cdot 9).\e$$

For $E_7$ we have, for any central charge not divisible by $2$ or $3$, 
$$D=2^m (k+18)^r,\e$$
where $m$ is the number of primary fields,
$$m=(k+1)(k+5)(k+7)(k+9)(k+11)(k+13)(k+17)/(6\cdot 8 \cdot 10\cdot  
12\cdot 14\cdot 18),\e$$
and 
$$r=(k-1)(k+1)(k+5)(k+7)(k+9)(k+11)(k+13) \cdot 7/
(9\cdot 2\cdot 6\cdot 8\cdot 10\cdot 12\cdot  14).\e$$

For $E_8$ we have, for any central charge, $k$, 
not divisible by $2,3,5$,
$$D=(k+30)^r.\e$$
Here, since $R=|M/M^*|=1$ (it is a self dual lattice), we have
only the power $r$, and $m$ is irrelevant. However, the 
same formula, eq. (35), holds
for the power $r$,
$$r=(k-1)(k+1)(k+7)(k+11)(k+13)(k+17)(k+19)(k+23)\cdot 
4/(15\cdot 2\cdot 8
\cdot 12\cdot 14\cdot 18\cdot 20\cdot 24).\e$$

We see that these are indeed polynomials of order ${\rm rank}(G)$,
in accordance with Conjecture (2a). 

It will be useful at this stage to work out some cases. So, take
$SU(4)$ or $A_3$, as an example. From conjecture (2b) we have
$$D=4^m (k+4)^r,\e$$
where 
$$m=(k+1)(k+2)(k+3)/6,\e$$
is the number of primary fields at the central charge $k$,
and $r$ is
$$r=(k-1)(k+1)(k+2)/2.\e$$
Substituting some low $k$'s we get,
for the discriminant, $D(k)$ where $k$ is the central charge, 
$D(1)=2^8$, $D(2)=2^{26} 3^6$, $D(3)=2^{40} 7^{20}$,
$D(4)=2^{205}$, $D(5)=2^{112} 3^{168}$, $D(6)=2^{308} 5^{140}$, 
$D(7)=2^{240} 11^{216}$, etc.
These numbers can be calculated directly from eq. (17). A sample
program is given in the appendix. Substituting into
eqs. (52,53), we get the exact same numbers. We see that the numbers
tend to grow very fast, for example, $D(7)$ has $297$ digits.
It is thus, rather striking, that these are integers which are
product of small primes, i.e., the divisors of the level, $N$.

Let us turn now to the non-simply laced case. Here we have the 
following conjecture.

{\rm \bf Conjecture (2c)}: For the non simply laced algebras $B_n$,
$C_n$, $F_4$ and $G_2$, the discriminant is given by
$$R^m (k+g)^r,\e$$
where $m$ is proportional to the number of primary fields in the
dual algebra, shifted by $h-g$, except for $C_n$ which will
be listed separately. Using Conjecture (1) we may write,
$$m=m_0 \prod_i {k+m_i+g-h\over 1+m_i},\e$$
where $m_0$ is the number of primary fields for $k=1+h-g$. The level $k$ is assumed to obey,
$$k+g-h\neq 0\mod \twidle g_i,\e$$
where $\twidle g_i$ are any of the Dual Coxeter numbers.
For $r$ we have a formula similar from before,
$$r=r_0 m (k+a)/(k+g-1),\e$$
where $a$ is some number and $r_0$ some normalization. An exception
for this formula is for $G_2$ which we will list explicitly below.

For $B_n$ we have, 
$$D=2^m (k+2 n-1)^r,\e$$
and in accordance with Conjecture (2c),
$$m=k(k+2)(k+4)\ldots (k+2 n-2) /
(6\cdot 8\cdot 10\cdot\ldots\cdot (2n)),\e$$
for all even $k$. For $r$ we have,
$$r=k(k+2)(k+4)\ldots (k+2 n-4) [k+(n-2)/2]/(4\cdot 6\cdot\ldots
\cdot (2 n-2)).\e$$

For $C_n$ we have
$$D=2^m (k+n+1)^r.\e$$
For $m$ we have the exceptional formula, where half of the
exponents `mutate':
$$m=k(k+1)\ldots (k+n-2) (k+n)/(n-1)!,\e$$
and $r$ is given by eq. (57),
$$r=k(k+1)\ldots (k+n-3)(k+n-2)^2/(n-1)!.\e$$
Since for $C_n$ the dual Coxeter numbers are $1$, these formulas
hold for all the central charges, $k$.

For $F_4$ we have, 
$$D=2^m (k+9)^r,\e$$
where $m$ is again the number or primary fields in the dual 
algebra, shifted by $h-g=3$.
$$m=(k-2)(k+2)(k+4)(k+8)/(8\cdot 18),\e$$
where $k-3\neq 0 \mod 2,3$, i.e., $k$ is an even number not 
divisible by $3$. For $r$ we find, again as an exception to
conjecture (2c),
$$r=(k+1)(k+2)^2(k+4)/72.\e$$
Here two exponents have been mutated.

For $G_2$ we find
$$D=3^m (k+4)^r,\e$$
where $m$ is given by conjecture (2c), 
$$m=(k-1)(k+3)/2,\e$$
and $k$ is any odd number. The formula for $r$ is exceptional,
$$r=k^2+1.\e$$

This formulae summarizes all the algebras, albeit, only
for central charges $k$ obeying $k+g-h\neq 0\mod\twidle g_i$. 
For other $k$, not covered by conjecture (2), the factorization, 
eq. (34), does not always hold. 
In some cases, it does hold with polynomial expressions for the
powers, but, this is not always the case.
We leave the determination of these cases to further work.
Of course, in accordance with theorem
(2), the discriminant is always an integer which is a product of 
primes dividing the level, and we checked this for all the algebras,
with small $k$ not covered by conjecture (2).

We presented here the rather intriguing algebraic and arithmetical
properties of the discriminant. We hope that this would lead
to an arithmetic approach to the classification of RCFT problem.
In particular, one may ask what values of the discriminant may
appear in RCFT. Since the discriminant is an integer which
is a product of primes dividing the level, this question is
of a particular significance.
\ack
The author thanks Y. Frishman and M. Karliner for a discussion.

\refout

\Appendix{}

{\bf An example program for calculating the discriminant.}
\def\chapterlabel{A.}

Below is a sample Mathematica program which calculates the
discriminant for $SU(4)$ level $k$, with an example of $k=6$.
The program uses eq. (17) and the explicit form of the $A_3$ root system.

Here the weight is $m f_1+n f_2+l f_3$, where $f_i$ are the three fundamental weights.
For $\rho$ we take, $\rho=f_1+f_2+f_3$. We take for the positive roots
in eq. (17) their explicit form, $a_1$, $a_2$, $a_3$, $a_1+a_2$, 
$a_2+a_3$ and $a_1+a_2+a_3$,
where $a_1,a_2,a_3$ are the three simple roots. To calculate eq. 
(17), we use
the fact that $f_i a_j=\delta_{i,j}$.

Here $SA4$ calculates the modular matrix for $SU(4)$, and
$FA4$ calculates the discriminant as a product of the row of
the modular matrix to the power $-2$.
\obeylines

SA4[m\_, n\_, l\_] := N[(Sin[ (m +
     n + l + 3) Pi/(k + 4)]Sin[Pi (m + n + 2)  /(k + 4)]Sin[Pi (n + l + 
                  2) /(k + 4)] Sin[Pi (m + 1)/(k + 4)] Sin[Pi (l + 1) /(k + 
                  4)] Sin[Pi (n + 1) /(k + 4)]), ss];

FA4 := Function[k, dx := 0; Do[If[n1 + n2 + n3 $\leq$ k, dx =
     N[dx + SA4[n1, n2, n3]$^2$, ss]], $\{$n1, 0, k$\}$, $\{$n2, 0, k$\}$, $\{$n3, 0, k$\}$];
    dd := 1; Do[If[n1 + n2 + n3 $\leq$ k, dd = N[dd  SA4[n1, n2, n3]/Sqrt[dx], 
      ss]], $\{$n1, 0, k$\}$, $\{$n2, 0, k$\}$, $\{$n3, 0, k$\}$]; N[dd$\hat{\,}$(-2), ss]]

k := 6; ss := 280;

FactorInteger[Round[FA4[k]]]

$\{\{2, 308\}, \{5, 140\}\}$

\bye